\documentclass[12pt]{article}
\usepackage{epsfig}

\title{Formation of antideuterons in heavy ion collisions}
\author{{B.L. Ioffe, I.A. Shushpanov and K.N. Zyablyuk}\\
{\it \normalsize Institute of Theoretical and Experimental Physics},\\
{\it \normalsize B.Cheremushkinskaya 25, 117218 Moscow, Russia},\\
{\tt \normalsize ioffe@vitep1.itep.ru, shushpan@heron.itep.ru, zyablyuk@heron.itep.ru}}
\date{}
\begin{document}

\maketitle

\newcommand{\be}{\begin{equation}}
\newcommand{\ee}{\end{equation}}
\newcommand{\bea}{\begin{eqnarray}}
\newcommand{\eea}{\end{eqnarray}}
\newcommand{\ve}{\varepsilon}

\begin{abstract}
 The antideuteron production rate at high-energy heavy ions collisions is calculated basing on the concept of
 $\bar{d}$ formation by antinucleons which move in the mean field of the fireball constituents (mainly pions).
 The explicit formula is presented for the coalescence parameter $B_2$ in terms of deuteron binding energy
 and fireball volume.
\end{abstract}

\bigskip

\centerline{PACS numbers:  25.75.-q, 25.75.Dw, 12.38.Mh}

\vspace{10mm}

1.{\it Introduction}. Recent measurements have reported the production of antideuterons in heavy ion collisions
 \cite{SPS}, \cite{RHIC}, \cite{AGS}. The theoretical description of this interesting effect is complicated,
 because the antideuterons are produced at intermediate stage of fireball evolution,  when the density of hadronic
 matter is rather low, but the particle collisions are still important. In other words they are produced at the "dense
 gas" stage of the fireball evolution.

 Here we present the theoretical picture of this stage and calculate the ${\bar d}$ production. The basic
 ideas of our approach are the following. The dominant mechanism of ${\bar d}$ production is the
 formation of antideuterons through the fusion reaction $\bar{p} + \bar{n} \to \bar{d}$. The fusion reaction is not
 possible if all participating particles are on mass shell. However, in the fireball at the "dense gas" stage of its
 evolution $\bar{p}$, $\bar{n}$, $\bar{d}$ are not on mass shell, since they interact with surrounding matter.
 The interaction with the fireball constituents leads to appearance of the mass shift and widths of all particles
 propagating in the medium (or width broadening for unstable ones), analogous to refraction and attenuation
 indices in case of photon propagation. The fusion reaction rate is strongly enhanced in comparison with the
 main process of ${\bar d}$ production in vacuum $\bar{p} + \bar{n} \to \bar{d}+\pi$.  Another important
 ingredient of the theoretical picture is the balance of the deuteron formation and desintegration rates. This
 balance is achieved because of large number of produced pions and high rate of $\pi+{\bar d}$ collisions
 leading to ${\bar d}$ desintegration. The balance does not imply a statistical equilibrium, but rather a stationary
 process, like a balance in the isotope concentrations in  a radiative chain. The formation rate
 $\bar{p} + \bar{n} \to \bar{d}$ vanishes, when ${\bar d}$ size increases, i.e.~its binding energy $\ve\to 0$.
 This fact explicitly manifests itself in our calculations. The previous theoretical investigations of the problem
 were performed in statistical models \cite{7,8}, in the model of (anti)nucleon sources in the fireball \cite{Mr}
 and in the Wigner function approach (see \cite{GFR, NK} and references therein).
 In all these approaches the interaction of nucleons, forming ${\bar d}$ (or $d$) with the fireball
 constituens as well as ${\bar d}$ ($d$) desintegration was not accounted
 (in \cite{7, 8} the results do not depend on $\ve$).

 According to the dominant coalescence mechanism it is convenient to characterize $\bar{d}$ production in
 heavy ion collisions by the coalescence parameter:
 \be
 \label{coal}
 B_2 \, =\,  E_{\bar{d}} \frac{d^3 N_{\bar{d}}}{d^3 p_{\bar d}}   \left(
 E_{\bar{p}} \frac{d^3 N_{\bar{p}}}{d^3 p_{\bar{p}}} \, E_{\bar{n}} \frac{d^3
 N_{\bar{n}}}{d^3 p_{\bar{n}}} \right)^{-1},
 \ee
 where we can put $d^3N_{\bar{p}}/d^3 p_{\bar{p}}=d^3N_{\bar{n}}/d^3p_{\bar n},~p_{\bar{p}}=
 p_{\bar{n}}=p_{\bar{d}}/2$. In what follows, we will consider only the central heavy ion collisions.

 2. {\it Theory.} Consider the "dense gas" stage of fireball evolution,  which follows after so-called "chemical
 freeze-out" stage \cite{BHS,SB}. Assume, that particle propagations at this stage  may be described classically
 using kinetic equations. We use the notation $q_i(x,p)$, $i = \bar{p}$, $\bar{n}$, $\bar{d}$, $\pi$ $\ldots$
 for the double densities in coordinate and momentum spaces, $n_i(x) = \int~ q_i(x,p) d^3 p$ are the densities.
 ($q_i(x,p)$ -- are Lorentz invariant.) Let us work in the c.m.~system of colliding ions. The kinetic equation for
 $q_{\bar{d}}(p_{\bar{d}}, x)$ reads:
 $$
 \frac{m_{\bar{d}}}{E_{\bar{d}}} \frac{\partial q_{\bar{d}}(p_{\bar{d}},x)}{\partial x_{\mu}}
 \, u_{\mu}^{\bar{d}} \,= \, \frac{\partial q_{\bar{d}}}{\partial t} \, +\, {\bf v}_{\bar{d}} \nabla
 q_{\bar{d}} \, =\, \int d^3 p_{\bar{p}}\, d^3 p_{\bar{n}} \, q_{\bar{p}}
 (p_{\bar{p}}) \, q_{\bar{n}} (p_{\bar{n}})\,  \sigma_{\bar{p} \bar{n}
 \to \bar{d}} \, v^{rel}_{\bar{p} \bar{n}} \, \times
 $$
 \be
 \label{ke}
 \times  \delta^3(p_{\bar{p}}  + p_{\bar{n}} - p_{\bar{d}}) \, - \,
 q_{\bar{d}}(p_{\bar{d}}) \left[ \, \int d^3 p_{\pi} \, q_{\pi} (p_{\pi}) \,
 \sigma_{\pi \bar{d}} \, v^{rel}_{\pi \bar{d}} \, +  \, \ldots  \, \right]
 \ee
 where $u_{\mu}^{\bar{d}} = (1, {\bf v}_{\bar{d}})/ \sqrt{1-v^2_{\bar{d}}}$ is the $\bar{d}$ 4-velocity, the ellipsis mean
 similar terms for collisions of $\bar{d}$ with other constituents of the fireball ($p,n$ etc.) and
 $v^{rel}_{\bar{p} \bar{n}}$, $v^{rel}_{\pi \bar{d}}$ are the differences of $\bar{p}$, $\bar{n}$ and $\pi, \bar{d}$
 velocities $v^{rel}_{\bar{p}\bar{n}}=| {\bf v}_{\bar{p}}-{\bf v}_{\bar{n}} | $ etc. The terms, when $\bar{d}$
 appears in the momentum interval $p_{\bar{d}} + \Delta p_{\bar{d}}$ due to elastic collisions are neglected.
 Necessary applicability condition of (2) is $\lambda=p^{-1}_i \ll d$, where $d$ is the mean distance between
 fireball constituents.

 The cross section  $\sigma_{\bar{p} \bar{n} \to \bar{d}} = \sigma_{pn \to d}$ is equal to: 
 \be
 \sigma_{pn \to d} = \frac{3}{4} \cdot \frac{\pi}{4} ~
 \frac{g^2}{E_p E_n E_d}~ \frac{1}{v^{rel}_{pn}}~ \delta(E_p + E_n  - E_d),
 \ee
 where $E_p,E_n,E_d$ are $p,n$ and $d$ total energies, 3/4 is the spin factor and $g$ is the coupling constant of
 low energy effective $pnd$ intraction (in the $d$ c.m.~system). The value of $g^2$ was found by Landau
 \cite{9} from the requirement of coincidence (at the deuteron pole) of the $pn$ scattering amplitude in effective
 theory with the amplitude in the Bethe-Peierls theory of the low-energy $pn$-scattering \cite{10}. In the limit of
 zero range of nuclear forces $g^2$ is
 \be
 g^2 \,= \,128 \pi\, m_N \sqrt{m_N \varepsilon},
 \ee
 where $m_N$ is the nucleon mass, $\varepsilon = 2.2\, {\rm MeV}$ is the deuteron binding energy.  The account
 of non-zero range $r_0$ increases $g^2$ by a factor of $(1-\sqrt{m_N \varepsilon} r_0)^{-1} \approx 1.6$ \cite{10,11}.

 The mass of the particle moving in medium is shifted being compared with its vacuum value. Similarly, due to
 interaction with medium constutuents, the width $\Gamma$ appears (or width broadening, if the particle has its
 proper width). The mass shift $\Delta m(E)$ and $\Gamma(E)$ are expressed through the forward scattering
 amplitude $f(E)$ of the particle on medium constituent (see \cite{12,13} and references therein).
 \be
 \Delta \, m(E) \, = \, - \, 2 \pi \, \frac{n}{m} \, {\rm Re} \, f (E)
 \ee
 \be
 \label{wd}
 \Gamma(E)\, =\, 4 \pi \,\frac{n}{m}{\rm Im}\, f(E)\, =\, \frac{np}{m}\, \sigma(E),
 \ee
 where $E$, $p$ and $m$ are particle energy, momentum and mass, $n$ is the density of the constutuent in
 medium. Eqs.~(5),(6) take place in the system, where constutuents are at rest. In case of moving constituents
 the corresponding Lorentz boost must be done. (By definition $\Delta m$ and $\Gamma$ are Lorentz invariant,
 for details see \cite{14}).

 Therefore, $\bar{p}$, $\bar{n}$ and $\bar{d}$ in the reaction $\bar{p} + \bar{n}  \to \bar{d}$ can be considered
 as Breit-Wigner resonances with varying masses distributed according to the Breit-Wigner formula. In process
 of the fireball expansion these Breit-Wigner resonances smoothly evolve to their stable counterparts.
 So we integrate the first term in the r.h.s.~of (2) after substituting (3) over the masses $m'$ of
 the Breit-Wigner resonances:
 \bea
 I & = & \int \,d m_{\bar p}' \,d m_{\bar n}' \, d m_{\bar d}'  \,
 {\Gamma_{\bar p}/2\pi \over (m_{\bar p}'-m_{\bar p})^2+\Gamma_{\bar p}^2/4}\,
 {\Gamma_{\bar n}/2\pi \over (m_{\bar n}'-m_{\bar n})^2+\Gamma_{\bar n}^2/4}\,
 {{\tilde\Gamma}_{\bar d}/2\pi \over (m_{\bar d}'-m_{\bar d})^2+{\tilde\Gamma}_{\bar d}^2/4} \nonumber \\
 & & \times \,{3\pi\over 16}\, {g^2\over E_{\bar d}'}\, \int
 {d^3p_{\bar p}\over E_{\bar p}'} \, {d^3p_{\bar n}\over E_{\bar n}'} \,
 q_{\bar p}(p_{\bar p})\,q_{\bar n}(p_{\bar n}) \,
 \delta^3(p_{\bar p}+p_{\bar n}-p_{\bar d}) \, \delta(E_{\bar p}'+E_{\bar n}'-E_{\bar d}')
\eea
where $E_{\bar p}'=\sqrt{p_{\bar p}^2+m_{\bar p}'{}^2}$ etc.
We assume that the widths $\Gamma\ll m$ are much smaller than the typical momenta
 in ${\bar p}$, ${\bar n}$ distributions. Than the distributions $q_{\bar p}(p_{\bar p})=q_{\bar n}(p_{\bar n})$
 can be taken out from the integral sign at the values $p_{\bar p}=p_{\bar n}=p_{\bar d}/2$.
  The result of calculation is given by
 \be
 \label{kint}
 I \, =\, \frac{3 \pi^2}{16 \, E_{\bar{d}}}\,  g^2 \, \sqrt{\frac{\Gamma_{\bar p}+\Gamma_{\bar n}+{\tilde\Gamma}_{\bar d}
   }{m_N}} \, q^2_{\bar{p}} (p_{\bar{p}})
 \ee
 (the mass difference $\Delta m=m_{\bar d}-m_{\bar p}-m_{\bar n} \sim 30\, {\rm MeV}$
 is small in comparison with the width $\Gamma\sim 300 \, {\rm MeV}$ and neglected in (\ref{kint})).
 Later we assume $\Gamma_{\bar p}=\Gamma_{\bar n}\equiv \Gamma$.
 ${\tilde \Gamma}_{\bar d}$ generally is not equal to the antideuteron width
 $\Gamma_{\bar d}\approx 2\Gamma\sim 600 \, {\rm MeV}$. The ${\bar p}{\bar n}$ system with
 ${\bar d}$ quantum numbers at high excitations will not evolve to ${\bar d}$ in the process of fireball
 expansion, but may decay in other ways. One may expect ${\tilde\Gamma}_{\bar d}<\Gamma_{\bar d}$.
 We shall keep the ratio $a\equiv {\tilde \Gamma}_{\bar d}/\Gamma_{\bar d}$ as free parameter
 in the calculations.  However, the results weakly depend on this ratio: the variation within the limits
 $0<a<1$ may change the coalescence parameter (\ref{coal}) by at most $\sqrt{2}$ times, but in real cases
 about $20\%$. This uncertainty is within accuracy of the whole method, estimated as $50\%$.

 The contributions of direct processes ${\bar p}+{\bar p}\to {\bar d} + \pi^-$, ${\bar n}+{\bar n}\to {\bar d} + \pi^+$
 and ${\bar p}+{\bar n}\to {\bar d} + \pi^0$ are small, alltogether about $20\%$ in comparison with (\ref{kint}).
 If these processes were essential, the coalescence parameter $B_2$ (\ref{coal}) would be meaningless,
 since the antideuteron distribution $d^3 N_{\bar d}/d^3p_{\bar d}$ is given by complicated
 integral over the antinucleon distributions in this case.

 At large $p_{{\bar p}\bot}$ the ${\bar p}$ spectrum decreases steeply, so the approximation
 $p_{\bar p}=p_{\bar d}/2$ becomes inaccurate which leads to an underestimation of $B_2$.
 The case of large $p_{{\bar p}\bot}>1\,{\rm GeV}$ is not considered here.

 Using (\ref{wd}) and performing Lorentz boost to heavy ion c.m.~frame, the term in  square brackets in
 (\ref{ke}) can be brought to the form $(m_d/E_{\bar{d}}) \, \Gamma_{\bar{d}}$, where
 \be
 \label{tw}
 \Gamma_{\bar{d}}\, =\, \sum_i  \int d^3 p_i \, q_i (p_i) \, v^{rel}_{i \bar{d}} \, \sigma_{i \bar{d}}(\bar{d}~\mbox{at rest})
 \ee

 Suppose, that the rate of antideuteron collisions with other constituents of the fireball resulting in antideuteron
 desintegration is much larger than the rate of fireball expansion. This happens at collisions of heavy nuclei at
 high energies, when the fireball size is large because of large number of produced pions per nucleon. In this
 case one may expect the balance: the first term in the r.h.s. of (2) is equal to the second one and
 \be
 \label{ddist}
 q_{\bar{d}}(p_{\bar{d}}) \, =\, \frac{I}{\Gamma_{\bar{d}} (m_{\bar d}/E_{\bar{d}})}\, =\, \frac{3 \pi^2}{32\, m_N} \,
 \sqrt{1+a\over 2\, \Gamma\, m_N} \, g^2\, q^2_{\bar{p}} (p_{\bar{p}})
 \ee
 The momentum distribution $d^3 N_{\bar{d}}/d^3 p_{\bar{d}}$ entering (1) is obtained from (\ref{ddist})
 by integration over the fireball volume
 \be
 \label{con}
 \frac{d^3 N_{\bar{d}}(p_{\bar{d}})}{d^3 p_{\bar{d}}} \, =\, \int d^3x\, q_{\bar{d}}(p_{\bar{d}},x)
 \ee

 Using (1), (\ref{ddist}), (\ref{con}) and (4) (with $r_0$ correction) we find for the  coalescence parameter
 \be
 \label{b20}
 B^{th}_2 \, = \, \frac{24 \pi^3}{E_{\bar{p}}} \times 1.6 \, \sqrt{(1+a)\,\varepsilon\over 2\, \Gamma}  \,
 \,{\int d^3x \, q^2_{\bar p}(p_{\bar p}, x)\over \left[ \int d^3x \, q_{\bar p}(p_{\bar p}, x) \right]^2}
 \ee
 Since the $x$-dependence of $ q_{\bar p}(p_{\bar p}, x)$ is not known, we replace (\ref{b20}) by:
 \be
 \label{b21}
 B^{th}_2\, =\, \frac{24 \pi^3}{E_{\bar{p}}} \times 1.6 \, \sqrt{(1+a)\,\varepsilon\over 2\, \Gamma} \,
 \frac{2}{V}\,\frac{\overline{n^2}_{\bar{p}}}{(\bar{n}_{\bar{p}})^2}
 \ee
 where $V$ is the fireball volume, $\bar{n}_p$ and $\overline{n^2}_p$ are the mean and mean square $\bar{p}$
 densitites in the fireball. (The coordinate dependence of $\sqrt{\Gamma}$ is neglected). $B^{th}_2$ is Lorenz
 invariant, as it should be. The volume $V$ may be understood as a mean value of the fireball volume at a stage,
 where, on one side, hadrons are already formed, i.e., mean distances between them are larger than the
 confinement radius $R_c \sim 1/m_{\rho} \sim (1/4)\,{\rm  fm}$, but on the other side, hadron interactions are still
 essential. The antinucleon  distributions $n_{\bar{p}}({\bf r}),n_{\bar{n}}({\bf r})$ inside the  fireball are
 nonuniform: at the dense gas stage and before it  the  antinucleons strongly annihilate in the internal part of the
 fireball and in much less extent in its  external layer of the thickness of order $\bar{p}(\bar{n})$ annihilation
 length $l_{ann}$ (this effect was considered in \cite{Mr}). For this reason $\overline{n^2}_p/\bar{n}^2_p$ may
 be remarkable larger than 1. For the same reason the antinucleons and antideuterons from the backside of the
 fireball (relative to the observer) are absorbed in the fireball and cannot reach the detector (see Fig.~1).
 Therefore, only one half of the fireball volume contributes to the number of registered ${\bar p}$, ${\bar n}$
 and ${\bar d}$. The corresponding factor approximately equal to 2 is accounted in (\ref{b21}).

\begin{figure}[tb]
 \hspace{50mm} \epsfig{file=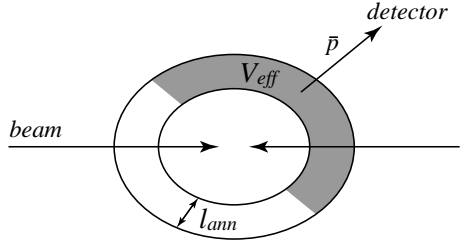, width=60mm}
 \caption{Fireball at the dense gas stage. The effective volume $V_{eff}$ is the half of the
 outer shell of thikness $l_{ann}$, from which the antiprotons reach detector}
 \label{fig_1}
 \end{figure}

 In fact, the fireball evolution after the balance may reduce the antideuteron number as
 $N_{\bar d} \to N_{\bar d} e^{-\Gamma_{\bar d}\Delta t}$, where $\Delta t$ is a typical time,
required for the antideuteron to leave the interaction region. (Such $\approx 50\%$  reduction of 
 $K^-$ mesons was observed in \cite{Pal}.) However, this effect does not change the 
coalescence parameter $B_2$. Indeed, in this case $B_2$ should be multiplied on the factor 
$e^{-\Gamma_{\bar d}\Delta t}/(e^{-\Gamma_{\bar p}\Delta t})^2$; the time $\Delta t$ is the same both for 
antideuteron and antiproton in consideration, because they move with equal velocities.
But since $\Gamma_{\bar d}=2\Gamma_{\bar p}$ with good accuracy (we checked it explicitly 
by eq (\ref{tw})), this factor is close to 1 regardless of the evolution details.

 $\Gamma$ may be calculated if the spectrum and densities of the fireball constituents at hadronic gas stage
 are known. At high energy of heavy ion collisions (SPS, RHIC) the main conributions to $\Gamma_{\bar p}$,
 $\Gamma_{\bar n}$, $\Gamma_{\bar d}$ come from the collisions of ${\bar p}$, ${\bar n}$, ${\bar d}$ with
 pions. Therefore $\Gamma_{\bar p}$,  $\Gamma_{\bar n}$, $\Gamma_{\bar d}$ essentially depend on
 pionic density in the fireball, the dependence of $\Gamma$ on the densities of other fireball constituents
 is much weaker. Also weak is the dependence of $\Gamma$ on the spectrum of the fireball constituents,
 since the main contribution to $\Gamma_{\bar p}$ ($\Gamma_{\bar n}$, $\Gamma_{\bar d}$) arises from
 the collisions at high energies in c.m.~system of ${\bar p}({\bar n}, {\bar d})+\pi$, where the cross sections
 are approximately constant. For this reason, without a serious error, for the calculation of
 $\Gamma$ we can take  the spectra from the experimental data, i.e.~corresponding to the final
 stage of the fireball evolution. Moreover, since the widths enters as $\sqrt{\Gamma}$ in (\ref{b21}),
 the errors are reduced twice. If $\Gamma$ is known, then by comparison with
 the data the parameter $V^{-1}(\overline{n^2}_{\bar{p}}/\bar{n}^2_{\bar{p}})$ can be found, what would allow to
 check various models of fireball evolution.

 3. {\it Comparison with the data}. Consider the NA44 experiment at SPS (CERN): $Pb+Pb$ collisions at
 $\sqrt{s} = 17 A \, {\rm GeV}$ \cite{SPS}. Antideuterons were observed at $0.6 < p_{\bar{d}t} < 1.6 \,{\rm GeV}$
 and in the rapidity interval 1.9 to 2.1 in lab.~system, which corresponds to $\bar{p}_{\bar{p}t} = 0.55\,{\rm GeV}$,
 $(E_{\bar{p}})_{c.m.} = 1.5\,{\rm GeV}$. The spectra and particle yields at such collisions are given in \cite{15}.
 The number of active nucleons, participating in collision ("wounded" nucleons) $N_N$ and the number of
 produced pions are presented in \cite{16}: $N_N = 362$, $N_{\pi} = 1890$, $Q_{\pi} = N_{\pi}/N_N = 5.2$
 (see also \cite{17} for the review of the data on heavy ion collisions).

 We accept the following model for the dense gas stage of fireball evolution \cite{13}. (A related model had
 been suggested long ago \cite{18,19}: it may be called Fermi--Pomeranhuk model). Neglect for a moment
 contributions of all particles except for nucleons and pions. Assume that  any participant -- nucleon or pion
 occupies the volume $v_N$ or $v_{\pi}$, respectively. Then
 \be
 \label{dens}
 n_N \,= \, \frac{N_N}{V}\, = \, \frac{n^0_N}{1 + Q_{\pi} \beta} \; , \qquad
 n_{\pi} \, =\, \frac{N_{\pi}}{V}\, =\, \frac{n^0_N Q_{\pi}}{1 + Q_{\pi} \beta}
 \ee
 where $n^0_N = 1/v_N$, $\beta = v_{\pi}/v_N$. For numerical estimations we take $n^0_N = 0.26 \, {\rm fm}^{-3}$,
 1.5 times standard nucleus density and $\beta = (r_{\pi}/r_N)^3 \approx 0.55$, where $r_{\pi} = 0.66\, {\rm  fm}$
 and $r_N = 0.81 \, {\rm fm}$ are pion and nucleon electric radii. It must be stressed, that $n^0_N$ is the only
 essential uncertain parameter in our approach. Even if $Q_\pi$ at dense gas stage differs from ones
 at the final stage, the arising error is essentially compensated by the appearance of $Q_\pi$ both in
 numerator and denominator in (\ref{dens}). (As was already mentioned, the nucleon contribution to
 $\Gamma$ is small.)

 Check first the applicability conditions of our approach. We have: $n = n_N + n_{\pi} \approx 0.42\, {\rm fm}^{-3}$
 and the mean distance between the fireball constutuents is $d = 1/n^{1/3} = 1.3\,{\rm fm}$. Evidently, the
 condition $\lambda_{\bar{p}} = 1/p_{\bar{p}} \ll d$ is well satisfied. The calculation of $\Gamma$ according to
 (\ref{tw}) ($\Gamma=\Gamma_{\bar d}/2$) gives $\Gamma\approx 300\,{\rm MeV}$. (Only inelastic cross
 sections were accounted, the pion contribution comprises about $75\%$, the nucleon one about $25\%$.
 Note, that the value of $\Gamma$ is close to the momentum integration interval in the Wigner function approach,
 $\Delta P\approx 200-300\,{\rm MeV}$, found in \cite{NK}.) Check now the balance condition -- that the
 probability of deuteron desintegration  exceeds the fireball expansion rate. The former is given by
 $2\Gamma(m_N/E_{\bar{p}})$. The estimation for the escape rate (or fireball expansion) is
 $w\sim (1/4)\,{\rm fm}^{-1}$. We have: $2 \Gamma(m_N/E_p) \approx 2.0\, {\rm fm}^{-1} \gg 0.25\,{\rm fm}^{-1}$.
 So, this condition is also fulfilled. Even more, the balance condition would be fullfilled at much lower hadronic
 densities than  chosen above, up to $n^0_N\approx 0.05\,{\rm fm}^3$, i.e. up to densities not much higher,
 than supposed for thermal freeze-out \cite{SB}, \cite{Ap}, \cite{St}. However, such low densities would
 lead to much lower values of $B_2$, than ones obtained in experiments.
 Eq.6 is legitimate, if $Im f(E) \ll d$ \cite{12,13}. Since $Im\, f \approx 1
 \,{\rm fm}$, this condition is not well satisfied. For this reason the value of $\Gamma$, presented above, has
 a large (may be 50\%) uncertainty, amd, probably, is overestimated (the effect of screening).
  This fact, however, does not influence too much the value $B^{th}_2$, since
 $\sqrt{\Gamma}$ enters (\ref{b20}). One may expect, that because of their slightly
  larger velocities in comparison with
 nucleons, pions form a  halo around  the fireball. This effect also may lead to an overestimation of $\Gamma$.

 At the parameters used above the fireball volume comes out to be:  $V = 6.2\times 10^3\,{\rm fm}^3$
 (15\% correction for other particles, except for pions and nucleons were accounted). This value is about 2 times
 larger, than the ones found in \cite{BHS} at chemical freeze-out and  about 2 times smaller, than at thermal
 freeze-out \cite{Ap,St}. (Note, that the dense gas stage is an intermediate between these two.) In the case of
 sphere its radius is equal to $R = 11.4\,{\rm fm}$. If we assume, that antiprotons  are mainly concentrated in the
 outer shell of the fireball  of the thickness of $l_{ann}\approx 3\,{\rm fm}$, then $\overline{n^2}/\bar{n}^2\approx 2$
 and we get for the coalescence parameter
 \be
 \label{b2sps}
 B^{th}_2\, =\, 3.0 \times 10^{-4}\,{\rm GeV}^2
 \ee
 (We put ${\tilde \Gamma}_{\bar d}=\Gamma$, or $a=1/2$.)
 Experimentally [1], for the average value of the most central 10\% events it was found: $B^{exp}_2 = (4.4 \pm 1.3)
 \times 10^{-4}\,{\rm GeV}^2$. However, $B^{exp}_2$ strongly depends on centrality: the results for $0-5\%$
 centrality are about 1.5 times lower. Taking in mind all uncertaintlies -- theoretical and experimental, we believe, that
 the NA44 data for coalescence parameter are not in contradiction with theoretical expectation.

 Turn now to the STAR experiment at RHIC: $Au+Au$ collisions at  $\sqrt{s} = 130 A \,{\rm GeV}$ \cite{RHIC}.
 Antideuterons were measured at $0.5 < p_t < 0.8\,{\rm GeV}$ and in the rapidity interval $\vert \Delta y_{c.m.}
 \vert < 0.3$, $18\%$ of central collisioins were collected.  We take  $\bar{E}_{\bar{p}, c.m.} = 1.05\,{\rm GeV}$.
 The number of "wounded" nucleons in the $18\%$ central $Au+Au$ collisions can be estimated as
 $N_N = 320$ \cite{20}. Multiplicity of negative hadrons $\bar{h}$ (mainly, pions)  was measured in \cite{21} at
 pseudorapidity $\eta = 0$ only  and it was found an increasing of $d h^-/d \eta\mid _{\eta=0}$ by $52\%$
 comparing with the SPS data at $\sqrt{s} = 17A\,{\rm GeV}$. But it is known that $d h/d \eta/_{\eta=0}$ increase
 faster with energy than the total multiplicity. We estimate $Q_{\pi} = N_{\pi}/N_N \approx 7 \pm 1$. (A value close
 to the presented above, can be found from the data compilation \cite{22}). At $N_N = 320$ with account of
 $20\%$ correction for $K$-mesons and hyperons  $V = 7.2\times 10^3\,{\rm fm}^3$. The coalescence parameter
 is equal to
 \be
 \label{b2st}
 B^{th}_2 = 3.8 \times 10^{-4}\, {\rm GeV}^2
 \ee
 ($\Gamma = 320\,{\rm MeV}$, $\bar{n}^2/(\bar{n})^2$ was put to be 2). Experimentally, STAR found $B^{exp}_2 =
 (4.5 \pm 0.3 \pm 1.0) \times 10^{-4}\,{\rm GeV}^2$.

 The main uncertainty of $B_2^{th}$ comes from the fireball volume $V$ which was calculated by (\ref{dens}).
 However, the width  $\Gamma$ also depends on the fireball volume, so that $B_2^{th}\sim 1/\sqrt{V}$, which
 suppresses this uncertainty twice. We expect the accuracy of our estimations (\ref{b2sps}), (\ref{b2st}) to be
 about  $50\%$.

 In  E864 experiment \cite{AGS} at AGS the antideuterons  were observed in  $Au+Pt$ collisions at
 $\sqrt{s}=4.8\,A\,{\rm GeV}$. 10\% of central collisions we selected. From the data we take: $p_{\bar{p}t} =
 0.17\,{\rm GeV}$, $\overline{E}_{\bar{p},c.m.}=0.99\,{\rm GeV}$. The number of "wounded'' nucleons and
 $\pi/N$ ratio are $N_N=350$, $Q_{\pi}=1.6$ (see \cite{17} and references herein). In the same way as before,
 we find: $V=2.8\times 10^3\,{\rm fm}^3$, $\Gamma=220\, {\rm MeV}$, $l_{ann}=1.2\,{\rm fm}$. In this case the
 validity conditions of our approach are at the edge of their applicability. So, the theoretical expectations for $B_2$
 are valid only by the order of magnitude:
 \be
 B^{th}_2 \sim 1.5 \times 10^{-3}\, {\rm GeV}^2
 \ee
 in comparison with $B^{exp}_2 = (4.1\pm 2.9\pm 2.3)\times 10^{-3}\,{\rm GeV}^2$.

 4.~{\it Summary and Acknowledgements.} The coalescence parameter $B_2$ for the antideuteron production in
 heavy ions collisions was calculated. It was supposed, that the $\bar{d}$ production proceeds at the stage, when
 the fireball may be treated as a dense gas of interacting hadrons. The $\bar{d}$ production is described as the
 formation process $\bar{p}+\bar{n}\to \bar{d}$, where $\bar{p}$, $\bar{n}$, $\bar{d}$ are moving in the mean field
 of the fireball constituents (mainly pions). It was shown, that in case of large $N_{\pi}/N_N$ ratio one may expect
 a balance: the number of produced antideuterons is equal to the number of desintegrated $\bar{d}$ due to
 collisions with pions. The balance  condition determines $\bar{d}$ production rate and the value of coalescence
 parameter $B_2$. The later is expressed in terms of deuteron binding energy and mean fireball volume at this
 stage. The comparison with data demonstrates, that ${\bar d}$-production proceeds at the stage intermediate
 between chemical and thermal freeze-out -- the dense gas stage of the fireball evolution. The theoretical values
 of $B_2$ are in satisfactory agreement with experimental data at SPS, RHIC and AGS but more data at various
 nuclei and various energies of collision and $\bar{d}$ energies would be very desirable. The comparison of the
 data with theory would allow to check various models of fireball evolution.

 We are thankful to G.Brown, L.McLerran, E.Shuryak for discussions and S.Kiselev, Yu.Kiselev, A.Smirnitsky,
 N.Rabin for information about experimental data. This work was supported in part by INTAS grant 2000-587
 and RFBR grant 03-02-16209.

\end{document}